\documentclass{emulateapj}

\usepackage{epsfig}
\usepackage{amssymb}
\shorttitle{Incidence Rate of GRB-host-DLAs}
\shortauthors{Nagamine et al.}

\begin{document}

\title{Incidence Rate of GRB-host-DLAs at High Redshift}

\author{Kentaro Nagamine\altaffilmark{1,2}, Bing Zhang\altaffilmark{1},
Lars Hernquist\altaffilmark{3}}
\altaffiltext{1}{University of Nevada Las Vegas, Department of Physics \& Astronomy, 4505 Maryland Pkwy, Box 454002, Las Vegas, NV 89154-4002 U.S.A.; Email: kn@physics.unlv.edu} 
\altaffiltext{2}{Visiting Researcher, Institute for the Physics and Mathematics of the Universe, 5-1-5 Kashiwanoha Kashiwa-shi, Chiba 277-8582 Japan}
\altaffiltext{3}{Harvard University, 60 Garden Street, Cambridge, MA 02138, U.S.A.}

\newcommand{\Mstar}{M_{\star}}
\newcommand{\Mbh}{M_{\rm BH}}
\newcommand{\Mhalo}{M_{\rm halo}}
\newcommand{\Mpeak}{M_{\rm peak}}
\newcommand{\Mmed}{M_{\rm med}}
\newcommand{\Mmean}{\avg{M_{\rm DLA}}}
\newcommand{\sdla}{\sigma_{\rm DLA}}
\newcommand{\Muv}{M_{\rm UV}}

\def\avg#1{\langle#1\rangle}
\newcommand{\Lam}{\Lambda}
\newcommand{\lam}{\lambda}
\newcommand{\Del}{\Delta}
\newcommand{\del}{\delta}
\newcommand{\mpc}{\rm Mpc}
\newcommand{\kpc}{\rm kpc}
\newcommand{\pc}{\rm pc}
\newcommand{\cm}{\rm cm}
\newcommand{\yr}{\rm yr}
\newcommand{\erg}{\rm erg}
\newcommand{\s}{\rm s}
\newcommand{\kms}{\,\rm km\, s^{-1}}
\newcommand{\Msun}{M_{\odot}}
\newcommand{\Lsun}{L_{\odot}}
\newcommand{\Zsun}{Z_{\odot}}
\newcommand{\hinv}{h^{-1}}
\newcommand{\himpc}{\hinv{\rm\,Mpc}}
\newcommand{\hikpc}{\hinv{\rm\,kpc}}
\newcommand{\himsun}{\,\hinv{\Msun}}

\newcommand{\Om}{\Omega_{\rm m}}
\newcommand{\Ol}{\Omega_{\Lam}}
\newcommand{\Ob}{\Omega_{\rm b}}
\newcommand{\OHI}{\Omega_{\rm HI}}
\newcommand{\HI}{H{\sc i}}
\newcommand{\NHI}{N_{\rm HI}}
\newcommand{\NHGRB}{N_{\rm HI}^{\rm GRB}}

\newcommand{\MgII}{Mg{\sc ii}}
\newcommand{\XH}{X_{\rm H}}
\newcommand{\Mtot}{M_{\rm tot}}
\newcommand{\Lbox}{L_{\rm box}}
\newcommand{\highz}{high-$z$}
\newcommand{\SFR}{{\rm SFR}}
\newcommand{\lgZ}{\log (Z/Z_\odot)}
\newcommand{\Ssfr}{\Sigma_{\rm SFR}}
\newcommand{\egrb}{\eta_{\rm GRB}}
\newcommand{\Lya}{{\rm Ly}\alpha}
\newcommand{\zgrb}{\zeta_{\rm GRB-host-DLA}}
\newcommand{\GDLA}{{\small GRB-host-DLA}}
\newcommand{\GDLAs}{{\small GRB-host-DLAs}}
\newcommand{\QDLA}{{\small QSO-DLA}}
\newcommand{\QDLAs}{{\small QSO-DLAs}}

%%%%%%%%%%%%%%%%%%%%%%%%%%%%%%%%%%%%%%%%%%%%%%%%%%%%%%%%%%%%%%%%%%%%%%

\begin{abstract}

We study the incidence rate of damped $\Lya$ systems associated with
the host galaxies of gamma-ray bursts (\GDLAs) as functions of neutral
hydrogen column density ($\NHI$) and projected star formation rate
(SFR) using cosmological SPH simulations.  Assuming that the
occurrence of GRBs is correlated with the local SFR, we find that the
median $\NHI$ of \GDLAs\ progressively shifts to lower $\NHI$ values
with increasing redshift, and the incidence rate of \GDLAs\ with $\log
\NHI > 21.0$ decreases rapidly at $z\ge 6$.
Our results suggest that the likelihood of observing the signature of
IGM attenuation in GRB afterglows increases towards higher redshift,
because it will not be blocked by the red damping wing of DLAs in the
GRB host galaxies.  This enhances the prospects of using high-redshift GRBs
to probe the reionization history of the Universe.  The overall
incidence rate of \GDLAs\ decreases monotonically with increasing
redshift, whereas that of \QDLAs\ increases up to $z=6$.  
A measurement of the difference between the two incidence rates would
enable an estimation of the value of $\egrb$, which is the mass
fraction of stars that become GRBs for a given amount of star formation.  
Our predictions can be tested by upcoming \highz\ GRB missions, 
including {\it JANUS (Joint Astrophysics Nascent Universe Scout)} 
and {\it SVOM (Space multi-band Variable Object Monitor)}. 
\end{abstract}


%%%%%%%%%%%%%%%%%%%%%%%%%%%%%%%%%%%%%%%%%%%%%%%%%%%%%%%%%%%%%%%%%%%%%%

\section{Introduction}
\label{section:intro}

A number of authors have proposed using GRBs to probe the history of
cosmic star formation and the reionization of the Universe
\citep[e.g.,][]{Totani97a, Miralda98a, Lamb00, Barkana04}, neither of
which is well-understood \citep[see, e.g.,][]{Holder03, Nag06c}.  
To date, observations of high-redshift (hereafter \highz) quasars and 
galaxies have been able to constrain reionization only up to $z\sim 7$ 
\citep{Fan06a}.
However, if GRBs are associated with the deaths of massive stars
\citep[e.g.,][]{Woosley93, Paczynski98}, then 
%the results from data analysis 
%\citep{Schaefer01} and theoretical modeling of early star formation 
%\citep{Abel02b, Bromm06, Yoshida07}
theoretical studies 
imply that GRBs may be detectable out to $z\simeq 10-20$ through their
prompt $\gamma$-ray emission and afterglows \citep{Lamb00, Ciardi00,
Gou04, Inoue07}.  
This raises the possibility of using GRBs to investigate the reionization 
history of the Universe, and the {\it Swift} satellite has indeed 
detected \highz\ GRBs with bright afterglows 
\citep{Cusumano06, Haislip06, Kawai06}.  

However, GRB lines-of-sight (LOSs) tend to probe more the inner parts 
of galaxies than the random QSO LOSs do, and are therefore often 
associated with neutral hydrogen (\HI) absorption \citep{Pro07a}.
Indeed, analyses of the afterglow spectra reveal the presence of DLAs 
in the red damping wing \citep{Vreeswijk04, Berger06, Watson06, Ruiz07}, 
and the DLAs may hide the absorption signatures of the neutral IGM 
\citep{Totani06}. 
If such cases dominate the \highz\ GRB afterglow spectra, then 
it may be difficult to use GRBs to probe the detailed reionization history 
of the Universe \citep{McQuinn08a}. Thus, it is important to understand 
the redshift evolution of the incidence rate of \GDLAs\ at $z\ge 6$ 
as a function of $\NHI$.

In this {\it Letter}, we use cosmological SPH simulations based on the
concordance $\Lam$ cold dark matter (CDM) model to study the $\NHI$
distribution and the incidence rate of \GDLAs\ as a function of
redshift between $z=1-10$.  The DLAs associated with quasar LOSs are
often referred to as \QDLAs.  Since quasars serve as randomly
distributed background beacons in the Universe, \QDLAs\ can be more
broadly interpreted as all the \HI\ gas clouds that satisfy the DLA
criterion ($\NHI > 2\times 10^{20}$\,cm$^{-2}$), regardless of whether
or not they have been intersected by quasar LOSs.  We adopt the latter
broad interpretation of \QDLAs\ in this paper, and by this definition
\GDLAs\ are a subset of \QDLAs.

\begin{figure*}
\begin{center}
\includegraphics[angle=0, scale=0.4]{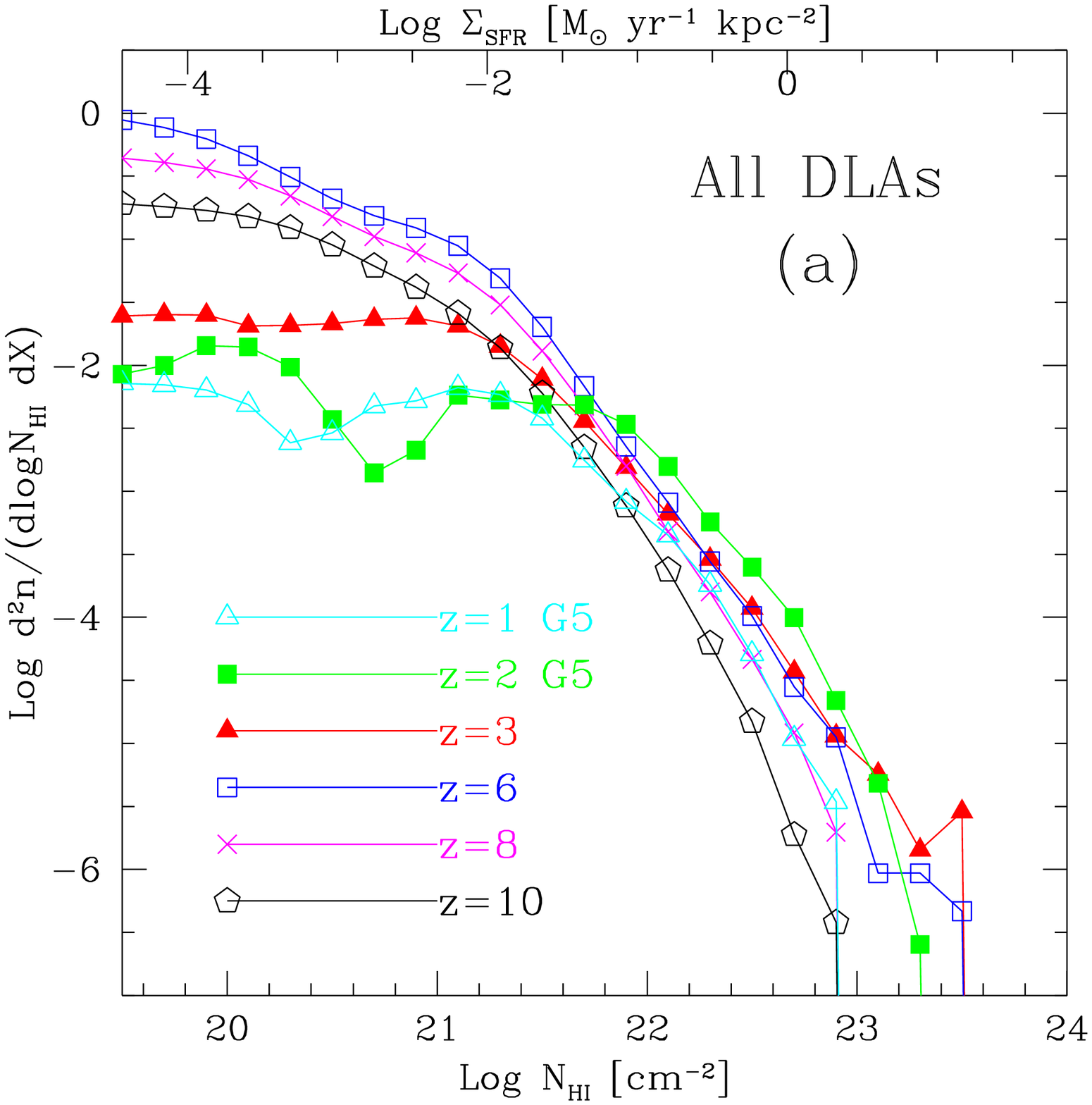}
\includegraphics[angle=0, scale=0.4]{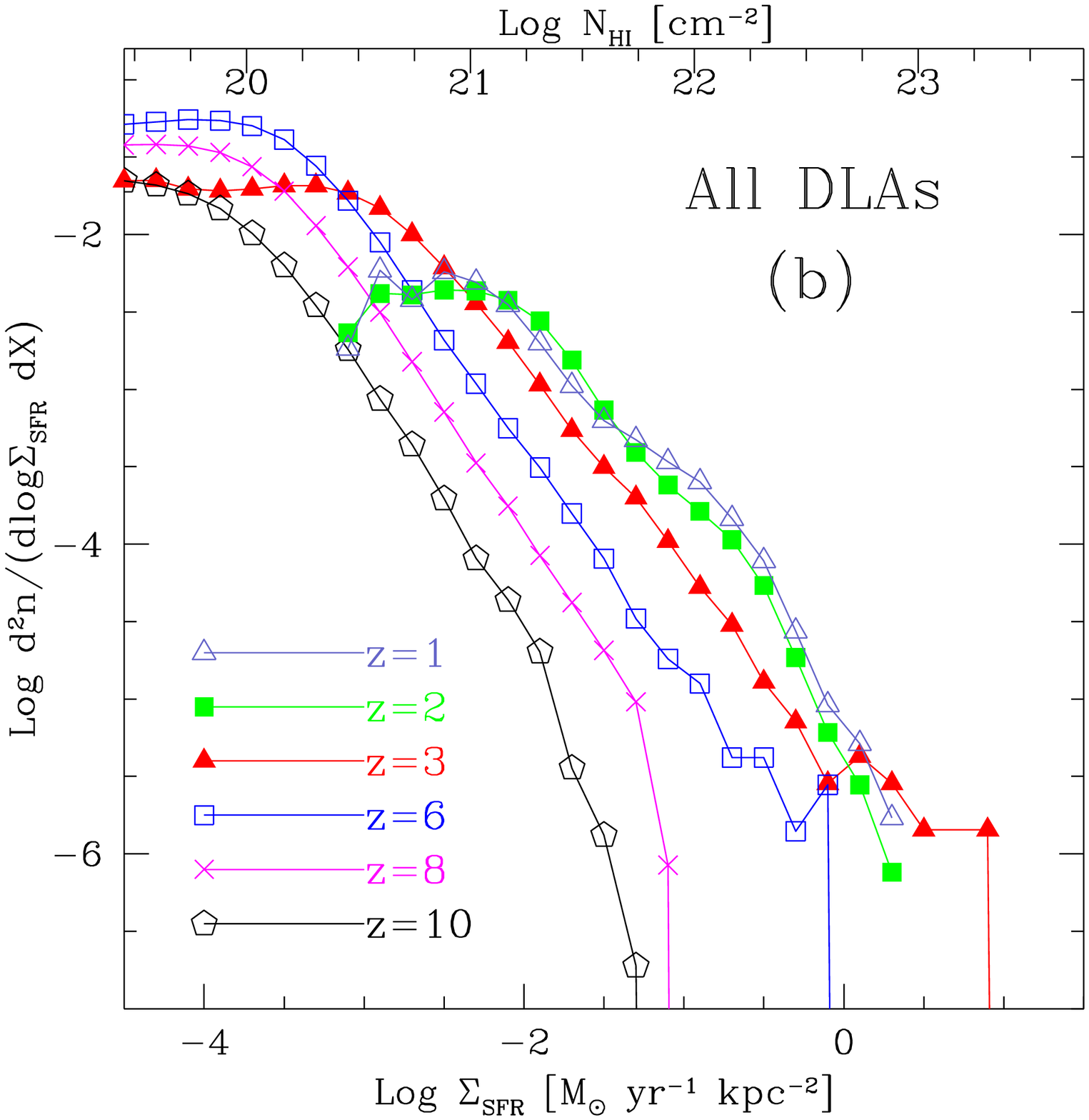}
\caption{Distribution of all DLAs (including both \QDLAs\ and \GDLAs)
as a function of $\NHI$ (panel [{\it a}]) and $\Ssfr$ (panel [{\it b}]) 
at $z=1-10$. 
The top axis in each panel indicates corresponding values of $\NHI$ and 
$\Ssfr$ based on the empirical \citet{Kennicutt98} law. 
}
\label{fig:nh}
\end{center}
\end{figure*}

%%%%%%%%%%%%%%%%%%%%%%%%%%%%%%%%%%%%%%%%%%%%%%%%%%%%%%%%%%%%%%%%%%%%%%

\section{Simulations}
\label{sec:sim}

Our simulations were performed with the smoothed particle hydrodynamics 
(SPH) code {\small GADGET-2} \citep{Springel05e}, which 
includes radiative cooling by hydrogen and helium, heating by a
uniform UV background \citep{Katz96a, Dave99}, star formation and
supernova feedback based on a sub-particle multiphase ISM model, and 
a phenomenological description of galactic winds
\citep{Springel03a, Springel03b}.

Here, we use the Q5 \& G5 runs from \citet{Springel03a}, which have box
sizes of 10 \& $100\,\himpc$, respectively.  The total particle number
is $N_p = 2\times 324^3$ for gas and dark matter in each run. The
initial gas particle mass is $m_{\rm gas}=3.3\times 10^5$ ($3.3\times
10^8$) $\himsun$, and the dark matter particle mass is $m_{\rm
dm}=2.1\times 10^6$ ($2.1\times 10^9$) $\himsun$. The comoving
gravitational softening length, a measure of the spatial resolution of 
the simulation, is $1.2$ (8.0) $\hikpc$ for the Q5 (G5) run.
We use the Q5 run for $z>3$ (for higher resolution), and the G5
run for $z<3$ (for better sampling of more massive halos and longer
wavelength perturbations).

Previously, we have used these simulations to study the properties of
DLAs \citep{Nag04g, Nag04f, Nag07a}, Ly-break galaxies at $z=3-6$
\citep{Nag04e, Night06}, and $\Lya$ emitters \citep{Nag08b}.  In
general, the simulations show reasonable agreement with available galaxy
observations, giving some confidence that we are capturing the basic
aspects of hierarchical galaxy evolution in the context of the
$\Lam$CDM model.  Moreover, the cosmic star formation history implied
by the simulations \citep{Her03} agrees with observational estimates
\citep{Faucher08a} and supports the association of GRBs with massive star
formation \citep{Faucher08b}.  The adopted cosmological parameters of all
simulations considered here are $(\Om,\Ol,\Ob,\sigma_8, h)= (0.3, 0.7,
0.04, 0.9, 0.7)$, where $h=H_0 / (100\kms\,\mpc^{-1})$.

%%%%%%%%%%%%%%%%%%%%%%%%%%%%%%%%%%%%%%%%%%%%%%%%%%%%%%%%%%%%%%%%%%%%%%

\section{Results}
\label{sec:NHI}

First, we present the distribution of all DLAs in the simulation as a
function of $\NHI$ in Figure~\ref{fig:nh}a.  The method of calculating
$\NHI$ in the simulations is described fully in \citet{Nag04g}.
Briefly, we set up a cubic grid that covers each dark matter halo, and
calculate $\NHI$ by projecting the \HI\ mass distribution onto one of
the planes.  The quantity ``$dn$'' is the area covering fraction on
the sky along the line element $cdt$ \citep[see Eqns.~5--7
of][]{Nag07a}, and the function $d^2n/(d\log \NHI\,dX) = f(\NHI, X)\,
\NHI \,\ln(10)$ is the `incidence rate' per unit $\log \NHI$ and per
unit absorption distance $X(z)$, where $dX=\frac{H_0}{H(z)}(1+z)^2
dz$.  The function $f(\NHI, X)$ is usually referred to as the column
density distribution function.

Figure~\ref{fig:nh}a shows that, from $z=10$ to $z=6$, the incidence
rate increases monotonically with decreasing redshift at all $\NHI$,
reflecting the rapidly growing number of dark matter halos.  From
$z=6$ to $z=3$, there is not much change at $\log\NHI>22$, but the
number of columns at $\log \NHI<21$ has decreased significantly, which
could owe to the UV background radiation field imposed at $z=6$ in the
simulation to model reionization.  From $z=3$ to $z=2$, the number of
columns at $\log\NHI>22$ increases, but it decreases from $z=2$ to
$z=1$, owing to the conversion of high-density gas into stars.

If long GRBs are associated with the collapse of massive stars, 
then their occurrence should be correlated with the local SFR. 
Therefore we define the following to quantify the distribution 
of \GDLAs:
\begin{equation}
\zgrb \equiv \frac{d^2 n}{dX d\log \Ssfr}~ \frac{\Ssfr}{\avg{\Ssfr}}~\egrb,
\label{eq:grb_rate} 
\end{equation}
where $\Ssfr$ is the projected SFR in units of $[\Msun\,\yr^{-1}\,\kpc^{-2}]$, 
and $\avg{\Ssfr}$ is a normalization parameter.  Although it is somewhat 
arbitrary, we take $\avg{\Ssfr}=10^{-4}$, because it roughly corresponds 
to $\log \NHI \approx 20$ based on the \citet{Kennicutt98} law.
The parameter $\egrb$ denotes the mass fraction of stars that become GRBs 
and have associated afterglows for a given amount of star formation 
with a certain stellar initial mass function (IMF).
The exact value of $\egrb$ depends on the IMF and other GRB physics.  
Here, we take $\egrb = 10^{-3}$, because the mass fraction of stars
with $M>8\,\Msun$ is 23\% of the total for a \citet{Chab03b} IMF with 
a mass range [0.1, 100]\,$\Msun$, and the global average of the
GRB/SN ratio is $\sim 0.5$\% \citep{Yoon06, Soderberg07, Campana08}.
If one takes the number fraction ($\sim 6$\% for $M>8\,\Msun$ stars) 
instead, then $\egrb = 3\times 10^{-4}$.  Here we use the mass fraction, 
because weighting by $\Ssfr/\avg{\Ssfr}$ is done on the basis of stellar mass. 

In principle, one could absorb the factor $\avg{\Ssfr}$ into $\egrb$
and treat them as one parameter: $\egrb^{\prime} \equiv \egrb/\avg{\Ssfr}$, 
which would be the GRB rate per projected SFR. 
However, here we choose to treat them separately to keep the 
physical meaning of $\egrb$ clear.  In the future, GRB theory 
may be able to estimate the value of $\egrb$, and observations of 
\GDLAs\ will constrain the ratio of $\egrb/\avg{\Ssfr}$
(see \S~\ref{sec:prob} and \ref{sec:discussion}). 

It is worthwhile to look at the distribution $d^2n/(dX d\log \Ssfr)$, 
before weighting it by $\Ssfr$.  Figure~\ref{fig:nh}b shows that the redshift 
evolution of this distribution is stronger than in Fig.~\ref{fig:nh}a.
The number of columns with $\log \Ssfr \gtrsim -2.5$ 
($\approx \log \NHI \gtrsim 21.0$ for the Kennicutt law) 
decreases systematically from $z=1$ to $z=10$.

\begin{figure}[t]
\begin{center}
\includegraphics[scale=0.4]{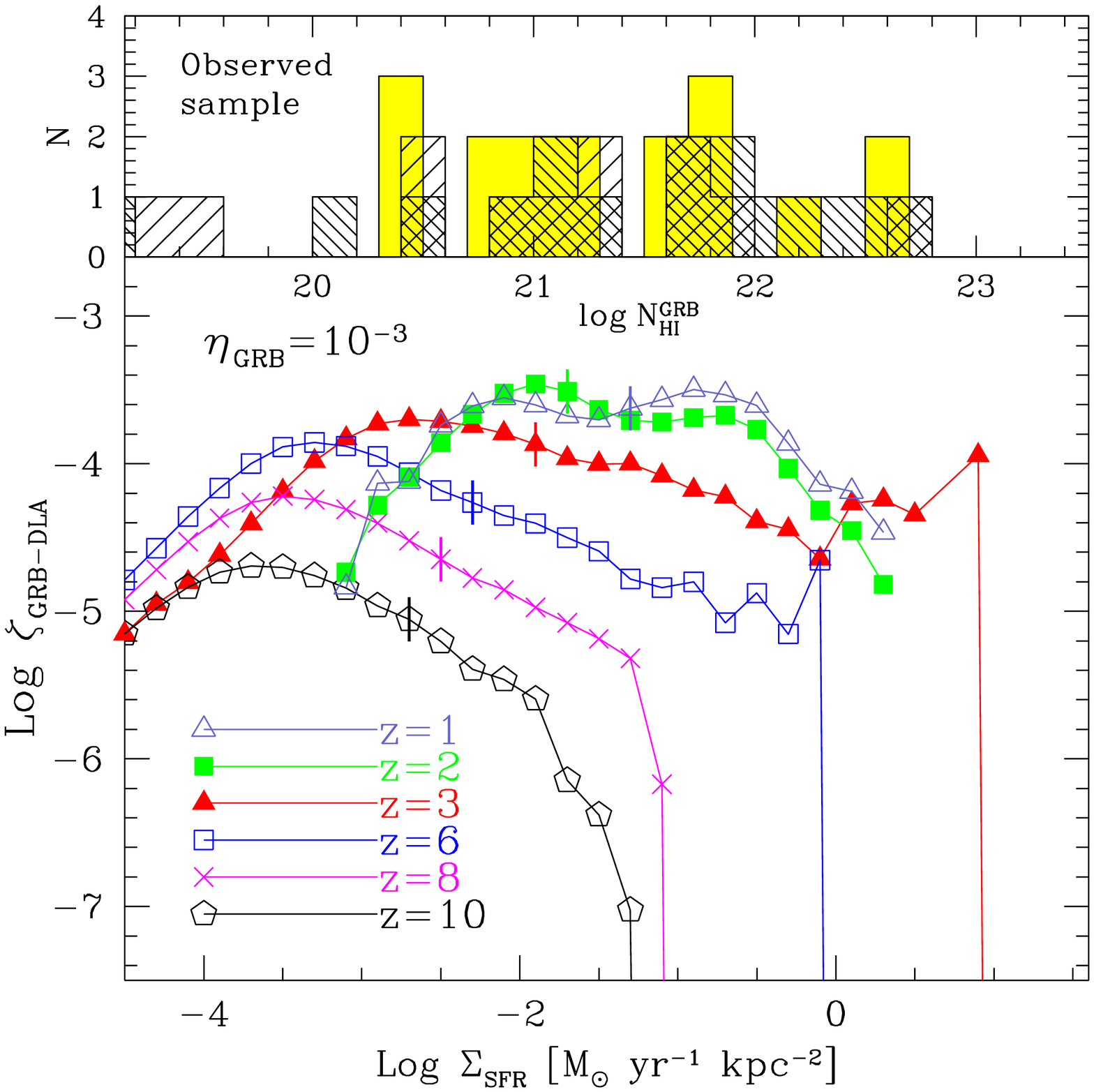}
\caption{
Distribution of \GDLAs\ as a function of $\Ssfr$ and $\NHGRB$ at $z=1-10$
is shown in the bottom panel.  The vertical tick-marks indicate the median 
of the distribution for columns with $\log \Ssfr > -3.0$ or 
$\log\NHGRB > 20.3$.  The top panel shows the observed sample of 28 \GDLAs\ 
from \citet[][broader (finer) hatching for $z<3$ ($z>3$)]{Chen07} and 
11 \GDLAs\ from \citet[][yellow shade; excludes those at $z<1.6$ with 
no $\NHI$ measurements]{Pro07a}.
The axis range of $\NHGRB$ corresponding to that of $\Ssfr$ is determined 
from the empirical \citet{Kennicutt98} law, divided by a factor of 2 
to take into account of the fact that the GRB LOSs go through only a half 
of the \HI\ slab on average compared to the QSO LOSs \citep{Pro07a}.
}
\label{fig:fsig}
\end{center}
\end{figure}

Figure~\ref{fig:fsig} shows $\zgrb$ as a function of $\log \Ssfr$. 
Because of the weighting by $\Ssfr/\avg{\Ssfr}$, the distribution now 
exhibits a peak with a broad tail at high $\Ssfr$. The number of columns 
with $\log \NHI > 21$ decreases progressively from $z=1$ to $z=10$,
owing to the decreasing number of massive dark matter halos with deep 
potential wells towards higher redshifts.  This is encouraging for the 
use of GRB afterglows to probe reionization, because the IGM attenuation 
signature is less likely to be blocked by the red damping wing of \GDLAs. 
The median value of the distribution at $\log \Ssfr > -3.0$ 
($\approx \log\NHI > 20.3$) is 
$\log \NHI = 21.4, 21.1, 21.0, 20.7, 20.6, \& 20.4$ for 
$z=1, 2, 3, 6, 8\, \& 10$, respectively. 

The distributions at $z=1$ \& 2 have broad peaks at $\log \NHI = 21.0-22.3$.
The current observed sample \citep{Chen07, Pro07a} is shown in the top
panel of Figure~\ref{fig:fsig}, and observationally there appears to be no 
indication that $\NHI$ drops towards \highz. If there is any observed trend, 
it might even be in the opposite sense, and all the observed \GDLAs\ are
at $z\gtrsim 2$. 
However the current observed sample is still small, and  observational 
selection effects are at play.  For example, systems with low $\NHI$ and 
low metal content are more difficult to identify, especially at higher 
redshift where the afterglows are typically fainter.
We also note that the decline of the distribution at $\log \Ssfr < -2.5$ 
for $z=1$ \& 2 may owe to the limited resolution of the G5 run compared 
to the Q5 run.

\begin{figure}[t]
\begin{center}
\includegraphics[scale=0.4]{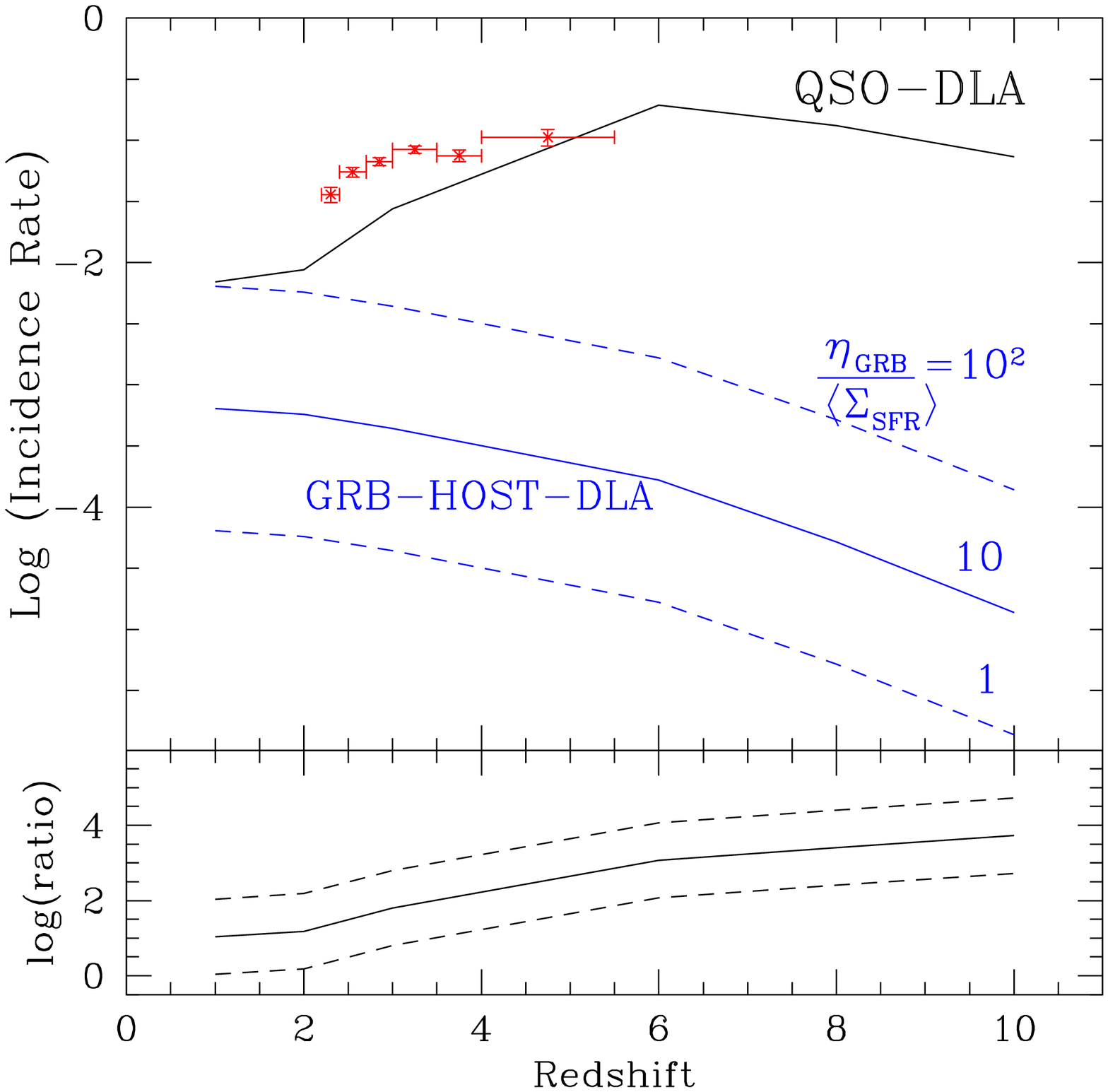}
\caption{{\it Top panel}: Incidence rates of \GDLAs\ and \QDLAs\ as 
a function of redshift.  (See text for the source of the data points.)
For \GDLAs, the three blue curves correspond to 
$\egrb/\avg{\Ssfr}=10^2, 10$, \& 1 from top to bottom.  
{\it Bottom panel}: Ratio of the two incidence rates.
}
\label{fig:rate}
\end{center}
\end{figure}

By integrating $\zgrb$ over $\log \Ssfr$, we obtain the `incidence rate' 
of \GDLAs.  The integral at $\log \Ssfr > -3.3$ yields rates of  
$(6.4, 5.7, 4.4, 1.7, 0.52, 0.14)\times 10^{-4}$ for $z=1, 2, 3, 6, 8\,\& 10$, 
respectively, for the assumed value of $\egrb / \avg{\Ssfr} = 10$. 
Figure~\ref{fig:rate} compares the derived incidence rate of \GDLAs\ to that 
of QSO-DLAs. The red data points are the updated
version\footnote{http://www.ucolick.org/\,$\tilde{}$\,xavier/SDSSDLA/} 
of \citet{Pro05} using SDSS DR5.  Our simulations somewhat underpredict the 
QSO-DLA incidence rate owing to the underestimate of $f(\NHI)$ at 
$\log \NHI < 21$ \citep{Nag04g}. 
There is a stark difference between the evolution of the two rates: 
the incidence rate of \GDLAs\ decreases monotonically towards \highz, 
whereas the QSO-DLA rate increases from $z=1$ to $z=6$. 
This is because we assumed that GRBs are correlated with star formation
and their distribution is not random, unlike the background quasars. 
The offset between the two rates tells us about the difference between 
the total \HI\ cross section of galaxies and the area covering fraction 
of star-forming regions.  The \QDLA\ sight-lines can also probe the 
outskirts of galaxies where star formation is nonexistent, 
therefore their incidence rate is much higher than that of \GDLAs.

%%%%%%%%%%%%%%%%%%%%%%%%%%%%%%%%%%%%%%%%%%%%%%%%%%%%%%%%%%%%%%%%%%%%%%

\section{Probability for a Given GRB}
\label{sec:prob}

Of immediate interest to GRB observers is the chance probability of 
finding a \GDLA\ for a given GRB event. 
For individual GRBs, the probability of having a \GDLA\ should depend 
on the geometry of the \HI\ gas distribution around the GRB along the 
LOS.  While our simulations do not have the resolution to 
follow the gas dynamics within molecular clouds, the Q5 run has a 
physical gravitational softening length of 0.31 (0.18)\,$\hikpc$ at 
$z=3$ (6), and the gas distribution above these scales is followed 
reasonably well.  The O and B stars would have ionized all the gas 
within $\sim 100$\,pc of the GRB creating an H{\sc ii} region, thus 
we expect the DLA gas to be located at $> 0.1$\,kpc from the GRB 
\citep{Pro07a}.  Then, our simulations can follow the qualitative trend 
in the redshift evolution of the incidence rate of \GDLAs. 

Since our simulations roughly match the Kennicutt law \citep{Nag04f}, 
it is guaranteed that columns with $\log \NHI > 20.3$ will have star 
formation along the LOS.  
Therefore if long GRBs are associated with star formation as we 
have assumed, most of the GRB LOSs will have a high-$\NHI$ gas 
in their host galaxy, majority of which are \GDLAs.
A more detailed analysis would be required to fully confirm this,
by generating the absorption line profiles for each GRB LOS, and
we plan to examine this in due course using higher resolution
simulations.

Nevertheless, Figure~\ref{fig:fsig} shows that the number of high-$\NHI$
systems decreases with increasing redshift, and so we expect that
the chance probability of having a high-$\NHI$ DLA for a given GRB event 
will also decline towards \highz\ with a similar qualitative trend as shown 
in Figure~\ref{fig:rate}.  But we stress that the incidence rate
shown in Figure~\ref{fig:rate} is {\it not} the probability of 
detecting a DLA for a given GRB. 

%%%%%%%%%%%%%%%%%%%%%%%%%%%%%%%%%%%%%%%%%%%%%%%%%%%%%%%%%%%%%%%%%%%%%%

\section{Conclusions \& Discussions}
\label{sec:discussion}

Using cosmological SPH simulations, we have examined the redshift 
evolution of incidence rates of \GDLAs, assuming that long GRBs are 
correlated with local SFR.
The distribution of \GDLAs\ is intrinsically different from that 
of \QDLAs, and the incidence rate of \GDLAs\ decreases monotonically 
towards \highz, whereas that of \QDLAs\ increases from 
$z=1$ to $z=6$. 
Quasars are assumed to be randomly distributed background sources 
in the sky, which illuminate the DLA gas in foreground galaxies. 
GRBs can also serve as randomly distributed beacons with respect to 
the DLA gas in foreground galaxies, but for \GDLAs, GRBs are not random 
background sources because they are in the same host galaxy.  

We find that the incidence rate of \GDLAs\ with $\log \NHI > 21.0$ 
decreases rapidly at $z\ge 6$, suggesting that the likelihood of 
observing the IGM attenuation signature in GRB afterglows increases 
toward higher redshifts, without being blocked by the red damping of 
DLAs in the GRB host galaxies.  This enhances the prospects for using 
\highz\ GRBs to probe the reionization history of the Universe.
Our predictions can be tested by upcoming \highz\ GRB missions, including
{\it JANUS (Joint Astrophysics Nascent Universe Scout)} and 
{\it SVOM (Space multi-band Variable Object Monitor)}. 

It might be hoped that it would be possible to estimate the incidence rate 
of \GDLAs\ by accumulating a large GRB sample.  However, because
GRBs are not random background sources for \GDLAs, this would require
a prohibitively large GRB sample to estimate the area covering fraction 
of \GDLAs\ from GRB observations alone.  If long GRBs do indeed trace 
star-forming regions, and if all the LOSs to star-forming regions are 
coincident with DLAs, then one could estimate the total \GDLA\ incidence 
rate simply by measuring the area covering fraction of star-forming 
regions from deep imaging surveys of galaxies. 
When the number of GRBs becomes comparable to that of QSOs, one should 
expect non-negligible new intervening DLAs in GRB LOSs that are not 
associated with GRB hosts, since GRB afterglows now act as random beacons.

An alternative possibility would be to search for quasars in the proximity 
of GRBs, or vice versa.  Such a search of QSO-GRB pair-LOSs would yield 
a coincidence probability between \QDLAs\ and \GDLAs, which roughly 
corresponds to the ratio of the two incidence rates.  
For this purpose, only those systems, for which the QSOs are in the 
background and the GRBs in the foreground, can be used.
So far, no such cases have been identified, but a combination 
of all-sky GRB surveys (e.g., BATSE, Swift, GLAST) and optical-IR imaging 
surveys of galaxies (e.g., SDSS, Pan-STARRS, LSST) may prove 
successful in the future. 
A constraint on the ratio of the two incidence rates would make it
possible to estimate the ratio $\egrb / \Ssfr$.
In addition, deep observations of GRB host galaxies could constrain $\Ssfr$
independently.  Then combining the above two constraints would allow us to 
estimate the value of $\egrb$.

%%%%%%%%%%%%%%%%%%%%%%%%%%%%%%%%%%%%%%%%%%%%%%%%%%%%%%%%%%%%%%%%%%%%%%

\section*{Acknowledgments}

This work is supported in part by the National Aeronautics 
and Space Administration under Grant/Cooperative Agreement No. NNX08AE57A 
issued by the Nevada NASA EPSCoR program and the President's 
Infrastructure Award from UNLV. 
BZ acknowledges the support from the NASA grant NNG06GH62G. 
We acknowledge the significant contribution of Volker Springel for the 
simulations used in this work. 
KN is grateful for the hospitality of Institute for the Physics and 
Mathematics of the Universe, The University of Tokyo, where part of 
this work was done. 
%The simulations were performed at the Institute of Theory and Computation at 
%Harvard-Smithsonian Center for Astrophysics, and the analyses were performed 
%at the UNLV Cosmology Computing Cluster.

%%%%%%%%%%%%%%%%%%%%%%%%%%%%%%%%%%%%%%%%%%%%%%%%%%%%%%%%%%%%%%%%%%%%%%

%\bibliographystyle{apj}
%\bibliography{ken}

\end{document}